\documentclass[12pt,preprintnumbers,aps,amssymb,nofootinbib,amsmath]{revtex4}
\usepackage{epsfig,epsf}
\usepackage{graphicx}
\usepackage{epstopdf}
\newcommand{\beq}{\begin{equation}}
\newcommand{\eeq}{\end{equation}}
\def\eq#1{{(\ref{#1})}}
\def\fig#1{{Fig.~\ref{#1}}}

\newcommand{\as}{\alpha_s}

\newcommand{\un}{\underline}
\newcommand{\ben}{\begin{eqnarray*}}
\newcommand{\een}{\end{eqnarray*}}
\begin{document} 

\preprint{RBRC-670}

\title{Forward hadron production \\ in high energy pA collisions: from RHIC to LHC }

\author{ Kirill Tuchin$^{a,b}$}
 
\affiliation{
a) Department of Physics and Astronomy,\\
Iowa State University, Ames, Iowa, 50011, USA\\
b) RIKEN BNL Research Center,\\
Upton, New York 11973-5000, USA\\
}

\date{\today}

\begin{abstract} 

We present a calculation of  inclusive $\pi$, $D$ and $B$ mesons production at RHIC and LHC energies based upon the KKT model of gluon saturation. We discuss dependence of the nuclear modification factor on rapidity, centrality and  transverse momentum.

\end{abstract}

\maketitle

%%%%%%%%%%%%%%%%%%%%%%%%%%%%%%%%%%%%%%%%

\vskip.3cm

\section{Introduction}\label{sec:intro}

In this paper we use the phenomenological model of Ref.~\cite{Kharzeev:2004yx} (KKT model) to analyze inclusive hadron production in pA collisions at RHIC and LHC. We focus on energy, rapidity, centrality and transverse momentum dependence of the nuclear modification factor $R_{pA}$ ($p$ stands for either proton or deuteron) for $\pi$, D and B meson production. Theoretical analysis \cite{Kharzeev:2002pc,Kharzeev:2003wz,Albacete:2003iq,Baier:2003hr} reveals that  coherent effects in hadron production become phenomenologically significant at RHIC. They consists of  (in the nucleus rest frame):  (i) multiple rescattering of the incoming parton system in the nucleus. This is controlled by the parameter $\as^2 A^{1/3}\sim 1$. And (ii)  small-$x$ quantum evolution controlled by the parameter $\as y=\as\ln(1/x) \sim 1$.  Both effects lead to the gluon saturation of a dense parton system at small $x$ which manifests itself in a different ways depending on the kinematic region we are interested in: (i) at not too small $x$ ($ e^{-1/\as}<x\ll 1$)  it contributes to the Cronin effect as the result of multiple rescatterings of the projectile in the target; (ii) at $x\lesssim e^{-1/\as}$  the small-$x$ evolution starts off 
which results in a dramatic suppression of $R_{pA}^\pi$ at forward rapidities at RHIC. Similar effects have been predicted \cite{Gelis:2003vh,Tuchin:2004rb,Blaizot:2004wv,Kovchegov:2006qn,Kharzeev:2003sk} and observed \cite{Averbeck} for open charm production. It is of great interest to observe the evolution of these phenomena with energy. Thus, motivated by the approaching start of the nuclear program at LHC we decided to combine calculations of inclusive hadron production in a framework of a single model. 

The paper is structured as follows. In Sec.~\ref{sec:model} we review a phenomenological model  for the dipole scattering amplitude $N(\un r,y)$ \cite{Kharzeev:2004yx} which has been previously used to successfully describe the data on light hadron production at RHIC. Although $N(\un r,y)$ can in principle be analytically calculated using the small-$x$ evolution equation \cite{Balitsky:1995ub,Kovchegov:1999yj}, its a very difficult and not yet solved problem. In addition, there is a fare amount of uncertainty in how the NLO terms in small-$x$ evolution change the behavior of $N(\un r,y)$ as well as  in the effect of fluctuations in the dense partonic system. On the other hand, rather general arguments allow identification of a few key features which hold for $N(\un r,y)$ in a dense parton system. This features are heeded in constructing a phenomenological models for $N(\un r,y)$. 
In this perspective we  argue in favor of the KKT model of Ref.~\cite{Kharzeev:2003wz}. 

In Sec.~\ref{sec:gluons} and Sec.~\ref{sec:quarks} we present theoretical results which together with the model of Sec.~\ref{sec:model} enables  calculation of pion and open charm and beauty production at RHIC and LHC. These results are summarized in figures \fig{fig1}, \fig{fig2} and  \fig{fig3}. They teach us that at LHC at rapidities $y\ge 0$, the nuclear modification factor is a very slow function of rapidity.  
This is an anticipated result. Indeed, the amount of suppression can be estimated as $R_{pA}\simeq 1/N_\mathrm{coll}^{1-\gamma}$, where $\gamma$ is a depends on rapidity and transverse momentum \cite{Kharzeev:2002pc,Kharzeev:2003wz}. 
Eqs.~ \eq{gamma} and \eq{xi}  imply that $\gamma$ decreases with energy and rapidity and increases with transverse momentum and mass, although it is a very slow function of its variables. Thus, we observe almost the same suppression pattern as the function of transverse momentum, rapidity and centrality for gluons and heavy quarks. To emphasize the mass and transverse momentum dependence of $R_{pA}$ we plotted it versus $m$ in \fig{fig4} up to the top-quark mass. Despite that $x$ is small in the kinematic region of this figure, the geometric scaling and hence the KKT model, are expected to break down. Still there is a vague theoretical understanding of how exactly it occurs. This too is discussed in  Sec.~\ref{sec:quarks}.

%%%%%%%
\section{A model} \label{sec:model}

In the dipole model \cite{dipole}, cross sections for inclusive hadron production can be expressed through the gluon dipole forward scattering amplitude $N(\un r,y)$, where $\un r$ is the transverse separation and $y$ is rapidity. This quantity can be calculated for any $\un r$ and $y>1$ using  the small-$x$ evolution equations, the most useful of which is the BK equation\cite{Balitsky:1995ub,Kovchegov:1999yj}.  In order to be able to utilize an exact numerical solution to BK \cite{Braun:2000wr,Lublinsky:2001yi,Armesto:2001fa,Levin:2001et,Golec-Biernat:2001if,Enberg:2005cb} one needs to (i) include the NLO corrections into the BK equation and (ii) perform global analysis of all small $x$ data.  Until this program is carried out, the most practical way to proceed  is to use a model for $N(\un r,y)$ which has 
the most essential properties of solution to the BK equation. Several such models have been suggested \cite{Golec-Biernat:1998js,Bartels:2002cj,Iancu:2003ge,Dumitru:2005kb,Dumitru:2005gt,Goncalves:2006yt,Kharzeev:2004yx}. Although they are not very much different from one another, the recent analysis of the numerical solution to BK equation performed in Ref.~\cite{Boer:2007wf} seems to imply that the KKT-like models are closer to the numerical solution than the other models. The main observation of Ref.~\cite{Boer:2007wf} is that the anomalous dimension $\gamma$ at the saturation scale is about 0.44. This should be compared to the anomalous dimension of 0.5 used by models based on the double logarithmic approximation, such as KKT, and to the anomalous dimension of 0.628 used by models based on the saddle-point approximation. It should be kept in mind that  $\gamma$  is the most important parameter determining the maximal possible suppression of the nuclear modification factor due to the gluon saturation effect. On the other hand,
the KKT model is certainly oversimplified and, as more accurate data appear, it will  be replaced by a more realistic models. 

For the reasons described above and encouraged by success of the KKT model in describing inclusive light hadron production at RHIC we set to employ it for calculation of inclusive hadron production at LHC. According to Ref.~\cite{Kharzeev:2004yx}  the forward gluon dipole scattering amplitude is parameterized as follows
\beq\label{modN}
N(\un r,y)=1-\exp\left\{ -\frac{1}{4}(r^2 Q_s^2)^{\gamma(r,y)}\right\}\,,
\eeq
where we denoted $|\un r|\equiv r$. The anomalous dimension is parameterized in such a way as to satisfy the analytically well-known limits of (i) $r\to 0$, $y$ fixed and (ii) $y\to \infty$, $r$ fixed:
\beq\label{gamma} 
\gamma(r,y)=\bigg\{ \begin{array}{ccc}
     \frac{1}{2}\left(1+ \frac{\xi(r,y)}{|\xi(r,y)|+\sqrt{2|\xi(r,y)|}+28\zeta(3)} \right) & y\ge y_0\,,\\
     1        & y<y_0\,,
     \end{array}
\eeq
where 
\beq\label{xii}
\xi(r,y)=\frac{\ln\left[ 1/(r^2 Q_{s0}^2)\right]}{(\lambda/2)(y-y_0)}\,.
\eeq
In the double logarithmic approximation we can replace 
$\un r^2\approx 1/(4 \un k^2)$.  The \emph{gluon} saturation scale  is given by
\beq\label{satt}
Q_s^2(y)=\Lambda^2\, A^{1/3}\, e^{\lambda y}=0.13\,\mathrm{GeV}^2\,e^{\lambda y}\,N_\mathrm{coll}\,.
\eeq
Parameters $\Lambda=0.6$ GeV and $\lambda=0.3$ are fixed by DIS data \cite{Golec-Biernat:1998js}. The minimal saturation scale used in \eq{xii} is defined by $Q_{s0}^2=Q_s^2(y_0)$ with $y_0$ the value of rapidity at which the small-$x$ quantum evolution effects set in. Fit to the RHIC data yields $y_0=0.5$ \cite{Kharzeev:2004yx} \footnote{In \cite{Kharzeev:2004yx} it has been argued that a non-perturbative scale (``intrinsic $k_T$") which is responsible for the Cronin effect at low energies has a little importance at RHIC. It is therefore  neglected in this paper.}. 

In the quasi-classical approximation (at $\gamma=1$) the forward scattering amplitude of the quark dipole $N_Q(\un r,y)$ is specified by the same formula \eq{modN} with the gluon saturation scale rescaled by the color factor $C_F/N_c$. Therefore, we model $N_Q(\un r,y)$ in the whole kinematic region by 
\beq\label{modNQ}
N_Q(\un r,y)=1-\exp\left\{ -\frac{1}{4}(r^2 Q_s^2/2)^{\gamma(r,y)}\right\}\,,
\eeq
where in the large $N_c$ limit $C_F/N_c=1/2$.

%%%%%%%%%%%%%%%%%%%%%%%%%%
\section{Light hadron production}\label{sec:gluons}

The light hadron production cross section is a sum of two terms:  inclusive gluon and valence quark production and hadronization. Inclusive gluon production dominates in most of the kinematic region safe for the most forward rapidities in the proton/deuteron fragmentation region, where 
Bjorken $x$ of  nucleus ($x_A$) acquires its lowest possible value for a
given $\sqrt{s}$, while that of  proton/deuteron ($x_p$) is close to
unity. In that region rescattering of valence quarks of proton in
 nucleus dominate to the hadron production
cross section. 

The cross section for inclusive gluon production is given by
\beq\label{sigma}
\frac{d\sigma_G}{d^2k\,dy}=\frac{\as C_F}{\pi^2}\frac{S_A}{k^2}
\, x_p^{-\lambda}(1-x_p)^4 \, \int_0^\infty dz_T \, J_0(k_T z_T) \, \ln\frac{1}{z_T\mu}
\, \partial_{z_T}[
z_T\,\partial_{z_T}N_G(z_T,y)],
\eeq
where $\mu$ is a scale associated with deuteron and is fixed at
$\mu=1$~GeV thereof. Expression $x_p^{-\lambda}(1-x_p)^4$ is a model for the gluon pdf of proton. Inclusive valence
quark production cross section \cite{jd}
\beq\label{jamal}
\frac{d\sigma_Q}{d^2k}=\frac{S_A}{2\pi}
\int_0^\infty dz_T \, z_T \, J_0(k_T z_T) \, [2 -  N_Q(z_T, y)],
\eeq

The nuclear modification factor is  defined as
\beq\label{rda}
R_{pA}(k,y)=\frac{\frac{d\sigma^h(pA)}{d^2k\,dy}}
{A\,\frac{d\sigma^h(pp)}{d^2k\,dy}}=
\frac{\frac{dN(pA)}{d^2k\,dy}}
{N_\mathrm{coll}\frac{dN(pp)}{d^2k\,dy}},
\eeq
where $\frac{dN(pA)}{d^2k\,dy}$ and $\frac{dN(pp)}{d^2k\,dy}$ are
multiplicities of hadrons per unit of phase space in pA and pp
collisions.  Both expressions for gluon \eq{sigma} and valence quark
\eq{jamal} production contribute to the hadron production cross section 
in pA and pp collisions. The cross section of hadron production reads
\begin{eqnarray}\label{hadr}
\frac{d\sigma^h}{d^2k\,dy}& =&
\int \frac{dz}{z^2} \, \frac{d\sigma_G}{d^2k\,dy}(k/z) \,
D_G (z, k) \, F(k/z,y) \nonumber\\ &&+\int
\frac{dz}{z^2} \, \frac{d\sigma_Q}{d^2k}(k/z) \,
xq_V(y,k/z)\, D_Q (z, k_T) \, F(k/z,y). 
\end{eqnarray}
We use the LO fragmentation functions from Ref.~\cite{frag} with
the renormalization scale of the fragmentation functions equal $k$.

Equation \eq{jamal} is derived for production of a valence quark in the 
proton/deuteron fragmentation region. To generalize it to smaller values of 
Bjorken $x$ one has to convolute it with the proton's/deuteron's valence quark 
distribution, which is fixed by quark counting rules at high $x$ and by the 
leading Regge trajectory at low $x$
\beq\label{qv}
xq_V(x)=1.09\,(1-x_p)^3\,x_p^{0.5}\,,
\eeq
where $x_p=(k_T/\sqrt{s})\,e^\eta$.  Valence quarks are increasingly
less important at low $x$ \cite{IKMT}, where the quark production is
dominated by gluons splitting in $q\bar q$ pairs. The factor of
$x_p^{0.5}$ insures that this is indeed the case here \cite{IKMT}.
Analogously, the high $x$ behavior of the \emph{nuclear} gluon
distribution is taken into account by introducing the function
$F(k,y)$
\beq\label{kinemat1}
F(k,y)=(1-x_A)^4\,
\left(\frac{\Lambda^2}{k^2+\Lambda^2}\right)^{1.3\,\as}\,.
\eeq
where the Bjorken $x$ of a gluon in the nuclear wave function is given
by $x_A=(k/\sqrt{s})\,e^{-\eta}$ and $\as =0.3$.  The last factor in
Eq.~(\ref{kinemat1}) arises when we impose momentum conservation
constraint on the anomalous dimension of the distribution
functions. 

Eq.~\eq{kinemat1} is an empiric way to include the higher order pQCD corrections. 
It is well-known that the NLO pQCD corrections are required to describe  the inclusive $\pi^0$ production in pp collisions at RHIC \cite{Jager:2002xm,Adler:2003pb}. Therefore, these corrections must be taken into account both in the numerator and the denominator of Eq.~\eq{rda}.  Unfortunately, the theoretical work on inclusion of the NLO corrections in the gluon saturation regime proved to be a very laborious problem. Only recently, calculations have been done for inclusion of some running coupling corrections into the forward scattering amplitude \cite{Gardi:2006rp,Kovchegov:2006vj,Balitsky:2006wa,Kovchegov:2006wf}.  This is the main reason why presently, we have to resort to the phenomenological approaches to the NLO corrections such as the one described here.

\begin{figure}[ht]
  \begin{tabular}{cc} 
      \includegraphics[width=8cm]{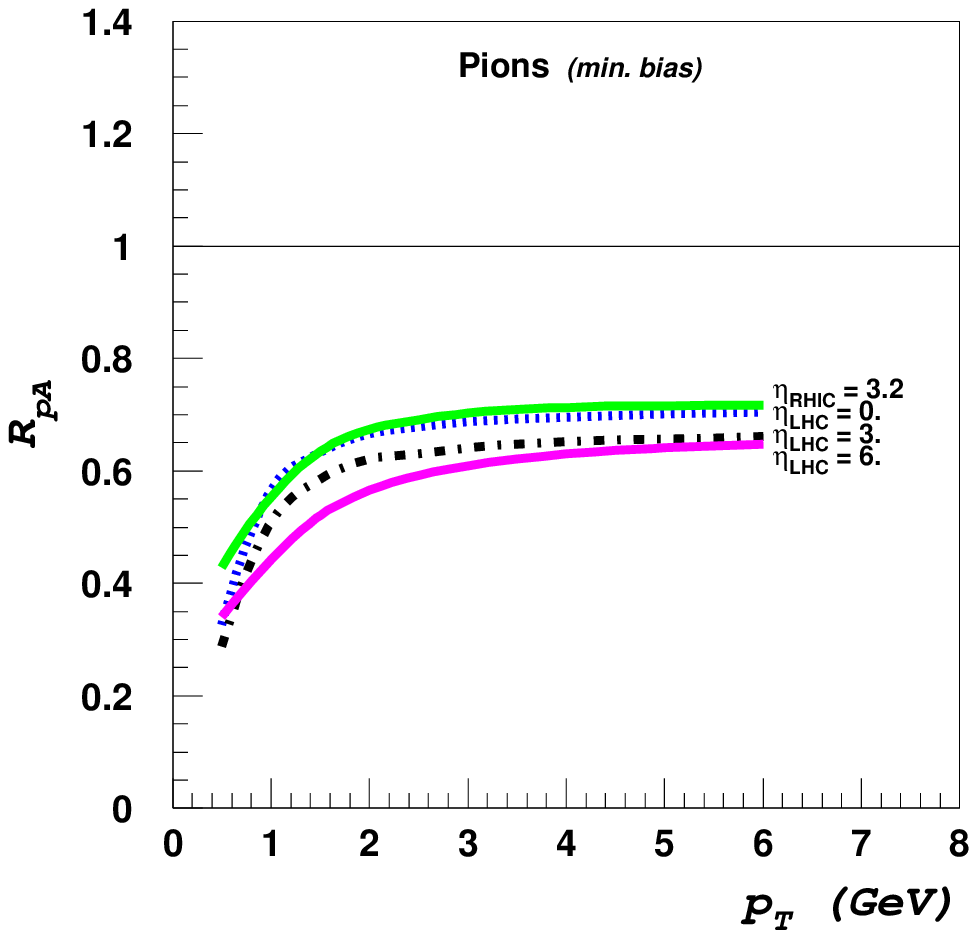} &
        \includegraphics[width=8cm]{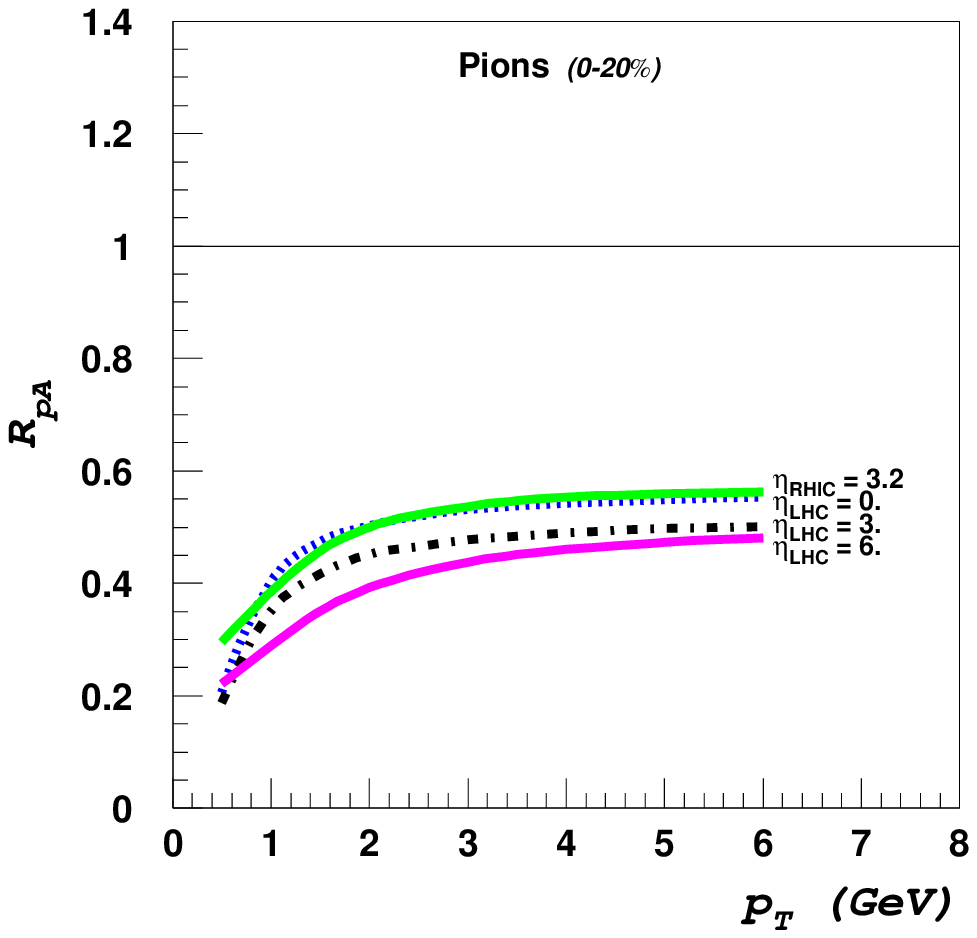}%\\
 %        (a) & (b) 
\end{tabular}
\caption{Nuclear modification factor for pion production at RHIC and LHC.}
\label{fig1}
\end{figure}

The results of numerical calculations are exhibited in \fig{fig1}. 
%%%%
%%%%%
As has been already pointed out in  \cite{Kharzeev:2004yx}, since rapidity  $\eta=3.2$ at RHIC corresponds to almost the same $x$ as $\eta=0$ at LHC, the nuclear modification factor is expected to be the same. This seen in \fig{fig1}. We  also observe that the difference between $\eta=0$ and $\eta =6$ is less than 20\%. 
This teaches us that fast onset of the gluon saturation observed at RHIC is replaced 
by virtually $y$-independent behavior at LHC (from $\eta=0$ to $\eta =6$). Behavior of $R_{pA}$ at rapidity $\eta_\mathrm{LHC}= -3.3$  and rapidity $\eta_\mathrm{RHIC}=0$ is expected to be similar. The amount of suppression perceived in \fig{fig1} depends mostly on centrality, i.e. $N_\mathrm{coll}$: $R_{pA}\simeq A^{-1/6}\simeq N_\mathrm{coll}^{-1/2}$. This yields $R_{pA}= 0.60$ for minimal bias events and  $R_{pA}=0.46$ for 0--20\% centrality cut in agreement with \fig{fig1}.

To conclude this section we would like to emphasize that we explicitly neglect a possible effect of gluon saturation in a proton or deuteron. This approximation  is \emph{perhaps} sufficiently  good for the nuclear modification factor which involves the ratio of the cross sections and is intentionally constructed to look for nuclear effects.

%%%%%%%%%%%%%%%%%%%%%%%%%%%%%%%
\section{Heavy quark production}\label{sec:quarks}

Production of heavy quarks at small $x$ is also affected by gluon saturation in a way similar to that of gluons \cite{Kharzeev:2003sk}. The main difference, however, is that the effect of gluon saturation is postponed to higher energies/rapidities  for heavier quarks as compared to lighter quarks and gluons. This is because the relevant $x$ is proportional to $m_\bot\sim(m^2+k_\bot^2)^{1/2}$ and hence is higher for heavier quarks at the same values of $\sqrt{s},y,k$. 

\begin{figure}[ht]
      \includegraphics[width=14cm]{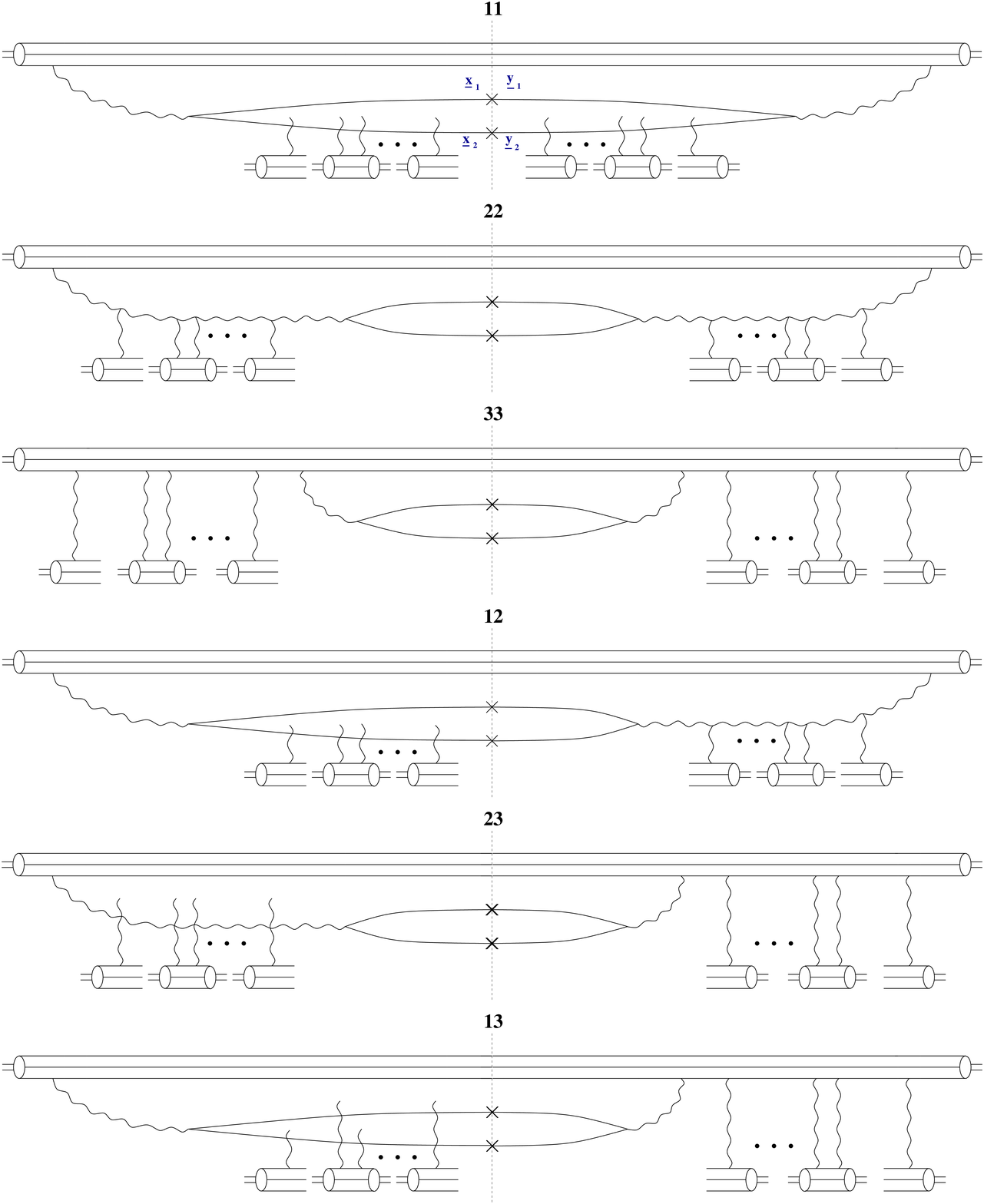} 
 \caption{The diagrams contributing to the quark--anti-quark pair production in the quasi-classical approximation. Disconnected $t$-channel gluon lines imply summation over all possible connections to the adjacent $s$-channel quark and gluon lines. Adopted from  \cite{Kovchegov:2006qn}.}
\label{fig:wave-functions}
\end{figure}

In this section,  we first of all review the theoretical approach to the $q\bar q$ pair production at small $x$, see Ref.~\cite{Tuchin:2004rb,Kovchegov:2006qn}. Let us introduce the following notations: $\un k$ is the produced quark transverse momentum, $\un q$ is the gluon transverse momentum, $\alpha=k_+/q_+$ is a fraction of the light-cone momentum of gluon carried by the produced quark; 
$\un x_1$ and $\un y_1$ are the transverse coordinates of the produced quark in the amplitude and in the complex conjugated amplitude respectively; $\un x_2$ and $\un y_2$ are the corresponding coordinates of the antiquark. Transverse coordinates of gluon in the amplitude $\un u$ and in the complex conjugated amplitude $\un v$ are given by  $\un
u \, \equiv \, \alpha \, \un x_1 + (1 - \alpha) \, \un x_2$ with $u =
|\un u|$ and $\un x_{12} = \un x_1 - \un x_2$, ($x_{12} = |\un
x_{12}|$) and analogously for $\un v$. With these notations 
the single inclusive quark production cross section is given by \cite{Tuchin:2004rb,Kovchegov:2006qn}
\beq
\frac{d \, \sigma}{d^2 k \, dy \, d^2 b} \,=\, \frac{1}{2 \, (2 \, \pi)^4} \, \int  
d^2 x_1 \, d^2 x_2 \, d^2 y_1 \, d^2y_2\,\delta(\un u-\un v-\alpha(\un x_{12}-\un y_{12}))\,\int_0^1 d\alpha \, e^{-i \, \un k \cdot (\un x_1-\un
y_1)}  \, \eeq
\beq\label{single_cl}
\times\,\sum_{i,j=1}^3\, \Phi_{ij}\, (\un x_1, \un x_2; \un y_1, \un
x_2; \alpha) 
\, \Xi_{ij} (\un x_1, \un x_2; \un y_1, \un x_2; \alpha)\,,
\eeq
where the delta function comes about after integration over the antiquark transverse momentum which implies $\un x_2=\un y_2$. 
The products of the light-cone ``wave functions" are detailed as follows \cite{Kovchegov:2006qn} see \fig{fig:wave-functions}
%%%%
%%%%%
 \begin{subequations}\label{phi}
\ben
\Phi_{11} (\un x_1, \un x_2; \un y_1,\un y_2 ; \alpha) \, = \, 4 \, C_F \, 
\bigg(\frac{\as}{\pi}\bigg)^2 \, \bigg\{ F_2(\un x_1, \un x_2; \alpha) \, 
F_2 (\un y_1, \un y_2; \alpha) \, \frac{1}{x_{12} \, y_{12} \, u \, v} \, 
[ (1 - 2 \, \alpha)^2 
\een
\ben
\times\,  (\un x_{12} \cdot \un u) \, (\un y_{12} \cdot \un v) 
+ (\epsilon_{ij} \, u_i \, x_{12 \, j}) \, (\epsilon_{kl} \, v_k \, y_{12 \, l})] +
F_1(\un x_1, \un x_2; \alpha) \, F_1(\un y_1, \un y_2; \alpha) \, m^2 \, 
\frac{\un u\cdot\un v}{u \, v} 
\een
\ben
+ 4 \, \alpha^2 \, (1-\alpha)^2 \,  F_0(\un x_1, \un x_2; \alpha) \, 
F_0(\un y_1, \un y_2; \alpha)  - 2 \, \alpha \, (1 - \alpha) \, (1 - 2 \, \alpha) \,
\bigg[ \frac{\un x_{12} \cdot \un u}{x_{12} \, u} \, F_2(\un x_1, \un x_2; \alpha) 
\een
\beq\label{phi11}
\times \,
F_0(\un y_1, \un y_2; \alpha) +  
\frac{\un y_{12} \cdot \un v}{y_{12} \, v} \, 
F_2(\un y_1, \un y_2; \alpha) \, F_0(\un x_1, \un x_2; \alpha) \bigg] \, \bigg\} \,,
\eeq
%%%%%%%%%%%%%%%%%%%%%%%%%%%%%%%%%%%%%%%%%%%%%%%%%%%%%%%
\ben
\Phi_{22} (\un x_1, \un x_2; \un y_1,\un y_2; \alpha) \, = \, 4 \, C_F \,
\bigg(\frac{\as}{\pi}\bigg)^2 \, m^2 \, \bigg\{
K_1(m \, x_{12}) \, K_1(m \, y_{12}) \, \frac{1}{x_{12} \, y_{12} \, u^2 \, v^2} 
\, [ (1 - 2 \, \alpha)^2
\een
\beq\label{phi22}
\times\,(\un x_{12} \cdot \un u) \, (\un y_{12} \cdot \un v) 
+ (\epsilon_{ij} \, u_i \, x_{12 \, j}) \, (\epsilon_{kl} \, v_k \, y_{12 \, l})] +
K_0(m \, x_{12})\, K_0(m \, y_{12}) \, \frac{\un u\cdot\un v}{u^2 \, v^{2}} \bigg\}\,,
\eeq
%%%%%%%%%%%%%%%%%%%%%%%%%%%%%%%%%%%%%%%%%%%%%%%%%%%%%%%%%%%%%%
\ben
\Phi_{12} (\un x_1, \un x_2; \un y_1, \un y_2; \alpha) = -  4 \, C_F \, 
\bigg(\frac{\as}{\pi}\bigg)^2 \, m \,
\bigg\{ F_2 (\un x_1, \un x_2;\alpha) \, K_1(m \, y_{12}) \, 
\frac{1}{x_{12} \, y_{12} \, u \, v^2} \, [ (1 - 2 \, \alpha)^2
\een
\ben
\times\, (\un x_{12} \cdot \un u) \, (\un y_{12} \cdot \un v) 
+ (\epsilon_{ij} \, u_i \, x_{12 \, j}) \, (\epsilon_{kl} \, v_k \, y_{12 \, l})]
+ m \, F_1(\un x_1,\un x_2; \alpha) 
\, K_0(m \, y_{12}) \, \frac{\un u\cdot\un v}{u \, v^{2}} 
\een
\beq\label{phi12}
- 2 \,  \alpha \, (1 - \alpha) \, (1 - 2 \, \alpha) \,
\frac{\un y_{12} \cdot \un v}{y_{12} \, v^2}  \, 
F_0(\un x_1, \un x_2; \alpha) \, K_1 (m \, y_{12}) \bigg\}\,, 
\eeq
%\end{eqnarray}
\ben
\Phi_{33} (\un x_1, \un x_2; \un y_1, \un y_2; \alpha) \, =\, 
\Phi_{11} (\un x_1, \un x_2; \un y_1, \un y_2; \alpha) + 
\Phi_{22} (\un x_1, \un x_2; \un y_1, \un y_2; \alpha) +
\Phi_{12} (\un x_1, \un x_2; \un y_1, \un y_2; \alpha) 
\een
\beq\label{phi33}
+ \Phi_{21} (\un x_1, \un x_2; \un y_1, \un y_2; \alpha)
\eeq
\beq\label{phi13}
\Phi_{13} (\un x_1, \un x_2; \un y_1, \un y_2; \alpha) \, =\, 
- \Phi_{11} (\un x_1, \un x_2; \un y_1, \un y_2; \alpha) 
- \Phi_{12} (\un x_1, \un x_2; \un y_1, \un y_2; \alpha)
\eeq
\beq\label{phi23}
\Phi_{23} (\un x_1, \un x_2; \un y_1, \un y_2; \alpha) \, =\, 
- \Phi_{21} (\un x_1, \un x_2; \un y_1, \un y_2; \alpha) 
- \Phi_{22} (\un x_1, \un x_2; \un y_1, \un y_2; \alpha)
\eeq
\beq\label{phi_symm}
\Phi_{ij} (\un x_1, \un x_2; \un y_1, \un y_2; \alpha) \, =\, 
\Phi_{ji}^* (\un y_1, \un y_2; \un x_1, \un x_2; \alpha).
\eeq
\end{subequations}
The auxiliary functions $F_1$, $F_2$ and $F_0$ are defined as 
\begin{subequations}
\begin{eqnarray}
F_2 (\un x_1, \un x_2; \alpha)&=& \int_0^\infty dq \, J_1 (q \, u) \,
K_1 \bigg(x_{12} \, \sqrt{m^2+
q^2\, \alpha(1-\alpha)} \bigg)\, \sqrt{m^2+
q^2\, \alpha(1-\alpha)} \,,\label{aux.fun1}\\ 
F_1 (\un x_1, \un x_2;\alpha)&=& \int_0^\infty dq \, J_1(q \, u) \, K_0 \bigg(x_{12}
\, \sqrt{m^2+q^2\, \alpha(1-\alpha)} \bigg)\,,\label{aux.fun2}\\
F_0 (\un x_1, \un x_2;\alpha)&=& \int_0^\infty dq \, q \, J_0(q \, u) \, K_0 \bigg(x_{12}
\, \sqrt{m^2+q^2\, \alpha(1-\alpha)} \bigg)\,,\label{aux.fun3}
\end{eqnarray}
\end{subequations}
where $u = |\un u|$, $\un x_{12} = \un x_1 - \un x_2$, $x_{12} =
|\un x_{12}|$, and $\un q = \un k_1 + \un k_2$. 

Using the definition of the \emph{gluon} saturation scale $Q_s$ (see \eq{satt})
\beq
Q_s^2 \, = \, 4 \, \pi \, \as^2  \, \rho \, T(\un b)
\eeq
with $\rho$ the nucleon number density in the nucleus and $T(\un b)$
the nuclear profile function, we write (in the large $N_c$ approximation)
\begin{subequations}\label{xi}
\begin{eqnarray}
\Xi_{11} (\un x_1, \un x_2; \un y_1, \un y_2; \alpha) 
&=&e^{-\frac{1}{8}\, (\un x_1 -\un y_1)^2 \, Q_s^2 \, \ln 
(1/|\un x_1 -\un y_1| \, \mu)
-\frac{1}{8}\, (\un x_2 -\un y_2)^2 \, Q_s^2 \, 
\ln (1/|\un x_2 -\un y_2| \, \mu)}\,,\\
\Xi_{22} (\un x_1, \un x_2; \un y_1, \un y_2; \alpha) 
&=&e^{-\frac{1}{4}\, (\un u-\un v)^2 \, Q_s^2 \, 
\ln (1/ |\un u -\un v| \, \mu)}\,,\\
\Xi_{33} (\un x_1, \un x_2; \un y_1, \un y_2; \alpha) &=& 1\,,\\
\Xi_{12} (\un x_1, \un x_2; \un y_1, \un y_2; \alpha) &=& e^{-\frac{1}{8}\, 
(\un x_1 -\un v)^2 \, Q_s^2 \, 
\ln (1/ |\un x_1 -\un v| \, \mu) -\frac{1}{8}\, (\un x_2 -\un v)^2 \, Q_s^2 \, 
\ln (1/ |\un x_2 -\un v| \, \mu)}\, ,\\
\Xi_{23} (\un x_1, \un x_2; \un y_1, \un y_2; \alpha) &=& e^{-\frac{1}{4} \, u^2 \, 
Q_s^2 \, \ln (1/u \, \mu)}\,,\\
\Xi_{13} (\un x_1, \un x_2; \un y_1, \un y_2; \alpha) &=& e^{-\frac{1}{8} \, x_1^2 \, 
Q_s^2 \, \ln (1/x_1 \mu) - \frac{1}{8} \, x_2^2 \, Q_s^2 \, \ln (1/x_2 \mu)}\,
\end{eqnarray}
\end{subequations} 
 All other $\Xi_{ij}$'s can be found from the components listed in
\eq{xi} using
\beq
\Xi_{ij} (\un x_1, \un x_2; \un y_1, \un y_2; \alpha) \, = \, 
\Xi_{ji} (\un y_1, \un y_2; \un x_1, \un x_2; \alpha)
\eeq
similar to \eq{phi_symm}. 

If the typical gluon momentum $\un q$ is much smaller than the produced quark mass, the above expressions can be  significantly simplified. Indeed, since $\alpha(1-\alpha)\leq 1/4$ we get
 \beq\label{approx}
  \un q^2\alpha(1-\alpha)\ll m^2\,,
  \eeq
   and the auxiliary functions read
\begin{subequations}
\begin{eqnarray}\label{aux.simpl}
F_2(\un x_1,\un x_2;\alpha)&=&K_1(x_{12}m)\,m\,u^{-1}\,,\\
F_1(\un x_1,\un x_2;\alpha)&=&K_0(x_{12}m)\,u^{-1}\,,\\
F_0(\un x_1,\un x_2;\alpha)&=&0\,.
\end{eqnarray}
\end{subequations}
In this approximation the only non-vanishing products of  ``wave functions" are given by
$$
\Phi_{11}(\un x_1, \un x_2; \un y_1, \un y_2; \alpha) \, =\, 
\Phi_{22}(\un x_1, \un x_2; \un y_1, \un y_2; \alpha) \, =\, -
\Phi_{12}(\un x_1, \un x_2; \un y_1, \un y_2; \alpha) \, 
$$
\ben
 = \, 4 \, C_F \, 
\bigg(\frac{\as}{\pi}\bigg)^2 \,m^2\, \bigg\{ K_1(x_{12}m)\, K_1(y_{12}m)\,\frac{1}{x_{12} \, y_{12} \, u^2 \, v^2} \, 
[ (1 - 2 \, \alpha)^2 \,(\un x_{12} \cdot \un u) \, (\un y_{12} \cdot \un v) 
\een
\ben
+ (\epsilon_{ij} \, u_i \, x_{12 \, j}) \, (\epsilon_{kl} \, v_k \, y_{12 \, l})] +
K_0(x_{12}m)\,K_0(y_{12}m)\,\frac{\un u\cdot\un v}{u^2 \, v^2} \bigg\}\,.
\een
Averaging over all directions of  gluon emission from the valence quark using 
$\langle \epsilon_{ij} \, u_i \, x_{12 \, j} \, \epsilon_{kl} \, v_k \, y_{12 \, l}\rangle = 
(1/2)\,u\,v\,\un x_{12}\cdot \un y_{12}$ we arrive at the well-known result \cite{Tuchin:2004rb,KopTar}
$$
\Phi_{11}(\un x_1, \un x_2; \un y_1, \un y_2; \alpha) = 
$$
\beq\label{phiT}
4 \, C_F \, 
\bigg(\frac{\as}{\pi}\bigg)^2 \,\frac{m^2}{uv}\,\bigg\{ \frac{\un x_{12}\cdot \un y_{12}}{x_{12}y_{12}}[(1-\alpha^2)+\alpha^2]\,K_1(x_{12}m)\, K_1(y_{12}m)+K_0(x_{12}m)K_0(y_{12}m)\bigg\}
\eeq

Approximation \eq{approx}, employed to derive \eq{phiT}, amounts to the $k_T$-factorization of the gluon distribution of proton (see \eq{main}).  It breaks down if $x$ is so small that proton can no longer be regarded as a dilute object: it acquires a hard  scale $\un q^2\sim Q_{sp}^2$ which increases with energy. However, in such a case even a more general formulas \eq{phi}, \eq{xi} which go beyond the $k_T$-factorization of proton by do not include the high density effects,  
become invalid too as they neglect the \emph{nonlinear} gluon evolution in proton.  Discussion of the gluon saturation in proton is beyond the scope of this paper.

It is convenient to express the scattering amplitude in terms of the vectors $\un x_{12}$, $\un y_{12}$, $\un u$ and $\un v$ and integrate in \eq{single_cl} over these variables. We have
\begin{subequations}
\begin{eqnarray}
\un x_1-\un y_1&=&\un x_{12}-\un y_{12}\,,\\
\un x_1-\un y_1 &=& 0\,,\\
\un x_1 -\un v&=& \un x_{12}-\alpha\,\un y_{12}\,,\\
\un x_2-\un v&=& -\alpha\, \un y_{12}\,,\\
\un y_1-\un u&=& -\alpha\,\un x_{12}+\un y_{12}\,,\\
\un y_2-\un u&=& -\alpha\, \un x_{12}\,.
\end{eqnarray}
\end{subequations}
Thus, sum over all rescattering factors including the signs of $\Phi_{ij}$'s is given by (we omit logarithms  $\ln(1/x\mu)$ for brevity)
$$
\Xi(\un x_{12},\un y_{12};\alpha)=e^{-\frac{1}{4}(\un x_{12}-\un y_{12})^2(Q_s^2/2)}+
e^{-\frac{1}{2}\alpha^2(\un x_{12}-\un y_{12})^2(Q_s^2/2)}
$$
\beq\label{sumX}
-
e^{-\frac{1}{4}(\un x_{12}-\alpha\,\un y_{12})^2(Q_s^2/2)}\,e^{
-\frac{1}{4}\alpha^2\,y^2(Q_s^2/2)}-
e^{-\frac{1}{4}(\alpha\,\un x_{12}-\un y_{12})^2(Q_s^2/2)}\,
e^{-\frac{1}{4}\alpha^2\,x_{12}^2(Q_s^2/2)}\,.
\eeq

Using the delta function in \eq{single_cl} to integrate over $\un u$ and integrating over $\un v$ in the leading logarithmic approximation we derive the final result
$$
\frac{d\sigma}{d^2k \,dy\, d^2b}=\frac{C_F\,\as^2\,m^2}{4 \pi^5}\int_0^1 d\alpha
\int d^2x_{12}d^2y_{12}\, e^{-i\un k\cdot(\un x_{12}-\un y_{12})}\,\ln(1/\mu|\un x_{12}-\un y_{12}|)
$$
\beq\label{main}
\times \bigg\{ \frac{\un x_{12}\cdot \un y_{12}}{x_{12}y_{12}}[(1-\alpha^2)+\alpha^2]\,K_1(x_{12}m)\, K_1(y_{12}m)+K_0(x_{12}m)K_0(y_{12}m)\bigg\}\, \Xi(\un x_{12},\un y_{12};\alpha)
\eeq

Before we turn to the numerical results, we would like to remark on the large $N_c$ corrections to \eq{main}. In our approach these corrections arise in the scattering amplitudes \eq{xi}. Exact expressions for those amplitudes in the quasi-classical approximation were derived in \cite{Tuchin:2004rb,Blaizot:2004wv}. Numerical uncertainty due to omission of the $\sim 1/N_c^2$ terms is expected to be of the order of 10\%. A detailed analysis performed in \cite{Fujii:2005vj} suggests that these corrections are even smaller. Another source of the large $N_c$ corrections is the high energy gluon evolution beyond the mean-field approximation. These corrections were studied in detail in \cite{Rummukainen:2003ns} and are shown to be at the 1\% level. Alluding to the poorly known NLO corrections as well as the accuracy of the present experimental data,  we think that neglecting the large $N_c$ corrections is justified.

\begin{figure}[ht]
  \begin{tabular}{cc} 
      \includegraphics[width=8cm]{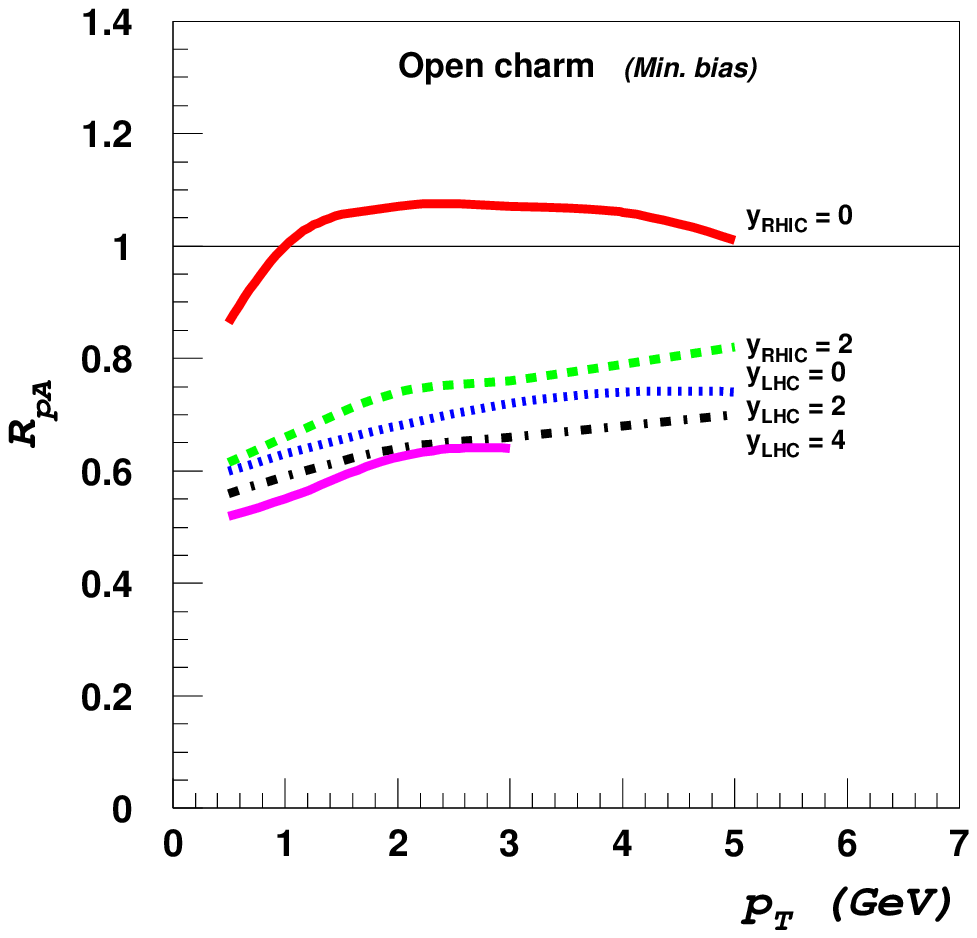} &
        \includegraphics[width=8cm]{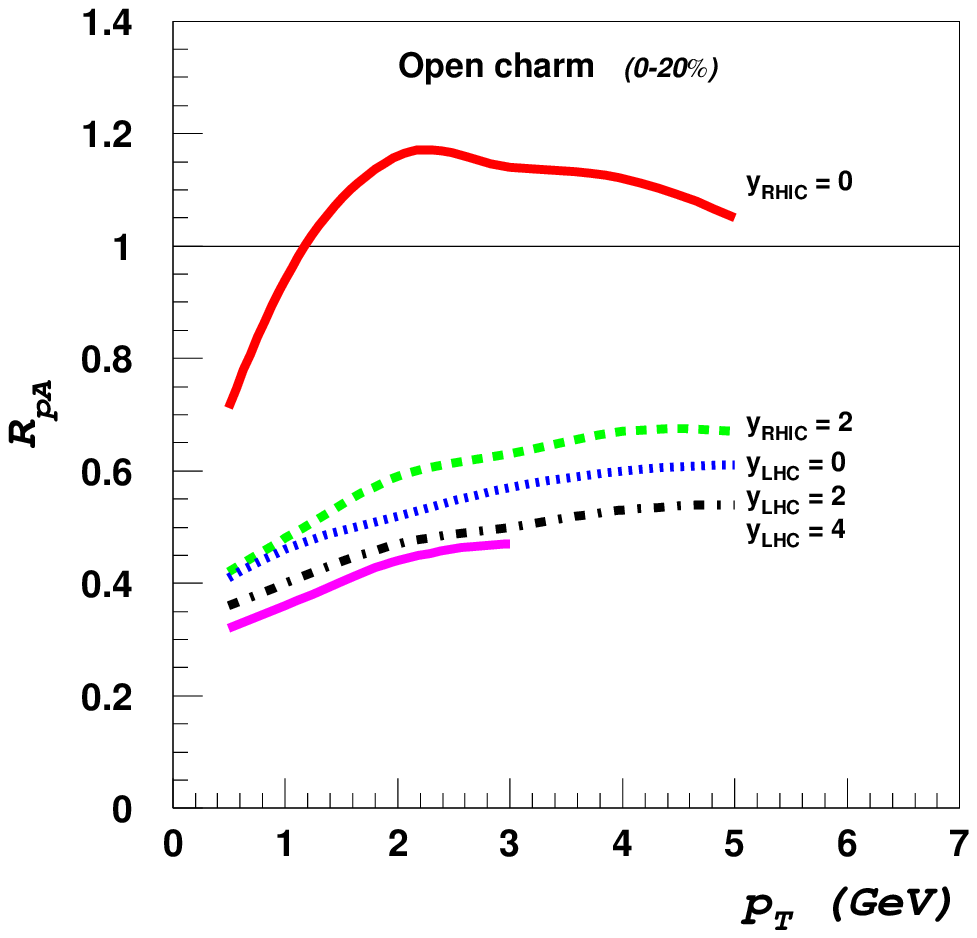}%\\
 %        (a) & (b) 
\end{tabular}
\caption{Nuclear modification factor for open charm production at RHIC and LHC.}
\label{fig2}
\end{figure}

\begin{figure}[ht]
  \begin{tabular}{cc} 
      \includegraphics[width=8cm]{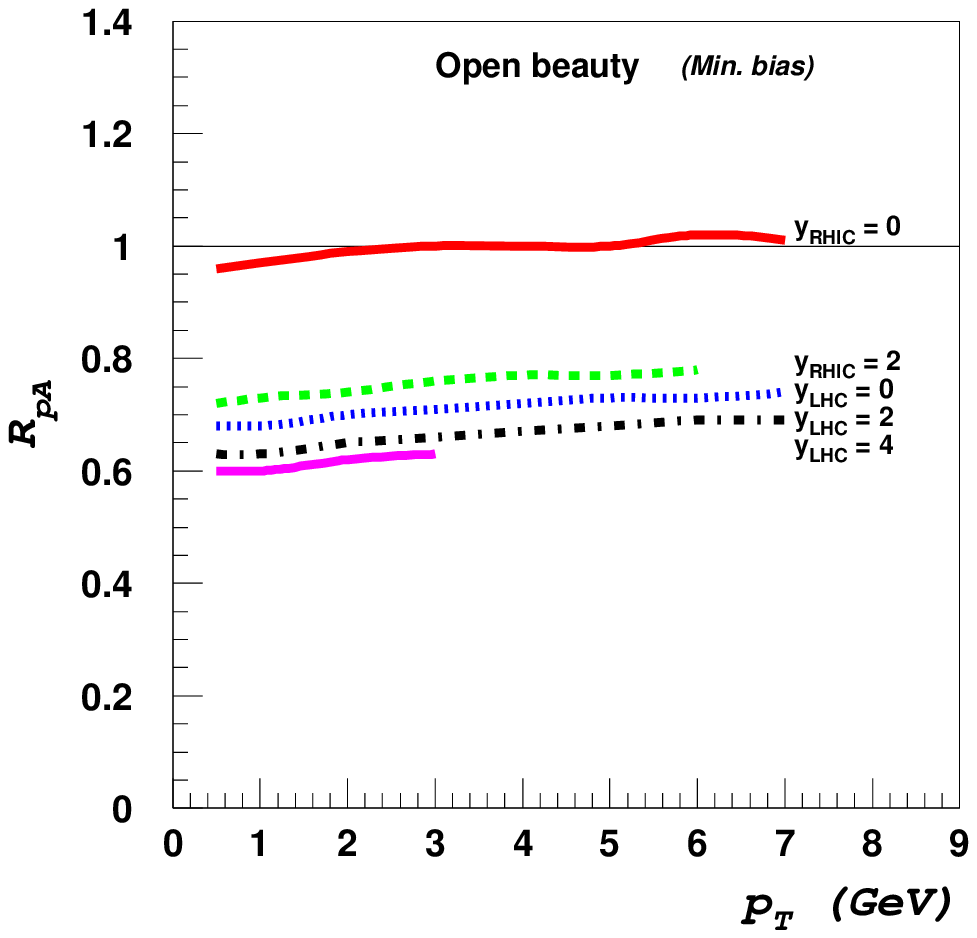} &
        \includegraphics[width=8cm]{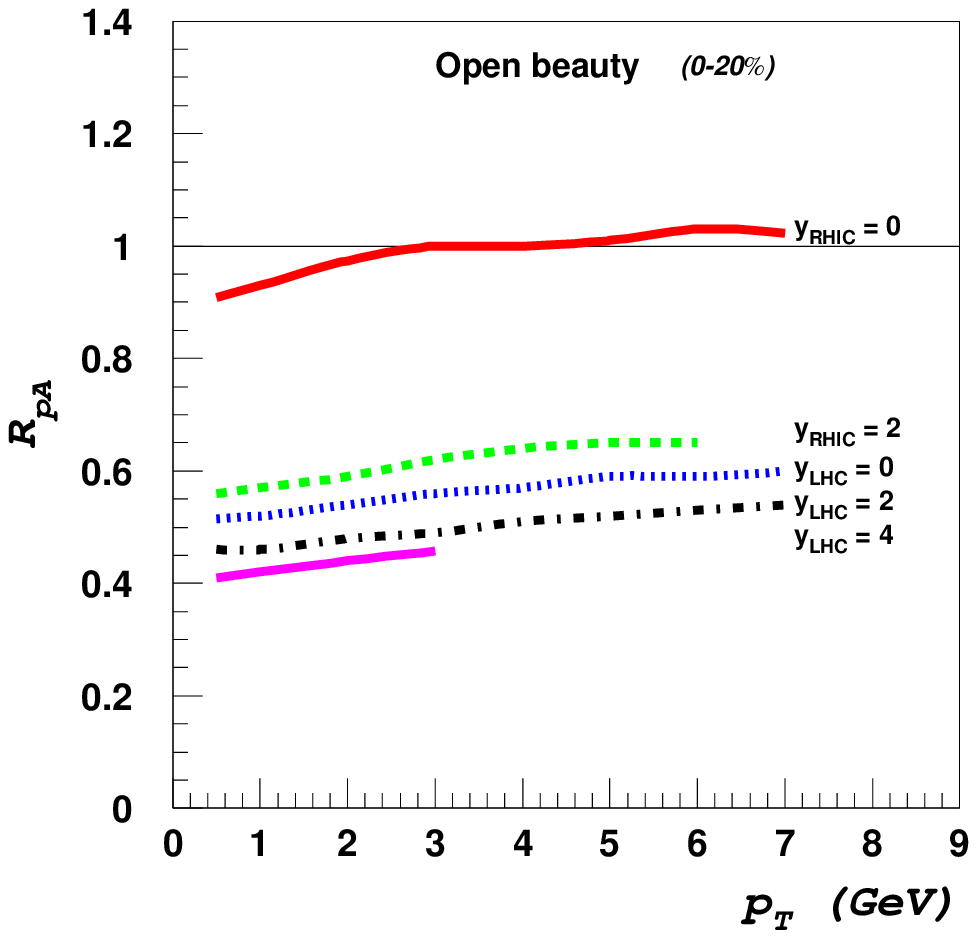}%\\
 %        (a) & (b) 
\end{tabular}
\caption{Nuclear modification factor for open beauty production at RHIC and LHC. }
\label{fig3}
\end{figure}

Numerical calculations are performed along the same steps as in Sec.~\ref{sec:gluons}: we replace the dipole scattering amplitude by the model \eq{modN} and \eq{modNQ}  and take into account the kinematic factors \eq{qv}, \eq{kinemat1}. 
In \eq{gamma} and \eq{xii} we substitute $\un r^2\approx 1/(4m_\bot^2)$. To obtain spectra of  D and B mesons we convolute the cross section \eq{main} with the Peterson fragmentation function $D(z)$ 
\beq\label{frag}
 \frac{d\sigma_\mathrm{hadron}}{d^2p \, dy}=
 \int_{z_\mathrm{min}}^1\,\int d^2k\, \frac{d\sigma(\un k, y)}{d^2k \, dy}\, \delta(\un p-z\un k)\, D(z)= \int_{z_\mathrm{min}}^1\frac{dz}{z^2}\,\frac{d\sigma(\un p/z,y)}{d^2k \, dy}\, D(z)\,,
\eeq
where $\un p$ is the hadron's transverse momentum and 
\beq
D(z)\propto \frac{1}{z}\left( 1-\frac{1}{z}-\frac{\epsilon}{1-z}\right)^{-2}\,,
\eeq
with $\epsilon =0.043$ for c-quark and $\epsilon =0.006$ for b-quark. 
The results of the calculations are displayed in \fig{fig2} and \fig{fig3}.

%%%%
%%%%%

%%%%
%%%%%

Note, that the calculations of \cite{Tuchin:2004rb,Kovchegov:2006qn} are applicable only if the coherence length of the typical $q\bar q$ pair is much larger than the size interaction of the interaction region which, in the nucleus rest frame, is about the size of the nucleus for the most central events. It is easy to verify (see e.\ g.\ \cite{Tuchin:2006hz}) that this condition holds for all lines in \fig{fig2} and \fig{fig3} apart from the B-meson production at $y=0$ at RHIC. In the former case, the hadron nuclear attenuation effect can be responsible for suppression (not show here) of B production. 

\begin{figure}[ht]
      \includegraphics[width=6.5cm]{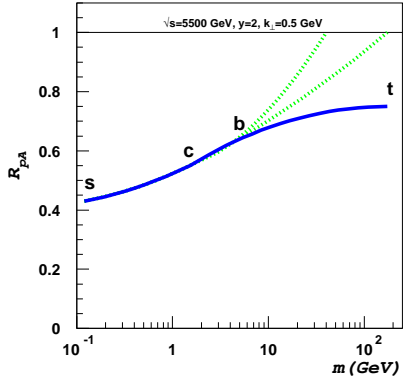} 
\caption{Dependence of the nuclear modification factor on quark mass. Solid line  is $R_{pA}$ for quarks (no fragmentation). Geometric scaling is expected to break down at $m_\bot\sim Q_\mathrm{geom}$ and therefore $R_{pA}$ is anticipated to deviate from the solid line towards unity. Dotted lines illustrates a possible behavior of $R_{pA}$. }
\label{fig4}
\end{figure}

In this paper we used a model for $N(\un r,y)$ which was tested in the RHIC kinematics. Extension to LHC involves an assumption that this model correctly  captures the energy dependence of the inclusive cross sections. It is of great interest to look in yet another kinematic region of low $x$ and high $m_\bot$. The very first paper on geometric scaling \cite{Stasto:2000er} hints that the geometric scaling holds in quite a wide kinematic region of  $x<0.01$ and $Q^2\le 450$~GeV$^2$ (and all small $x$ data available).
It was explained in \cite{Levin:2000mv} that this property stems 
directly from solution to BK equation. The geometric scaling is predicted to hold up to  momenta $Q_\mathrm{geom}\sim Q_s^2/\Lambda$ \cite{Iancu:2002tr}. Although there is a fair amount of ambiguity in the value of $\Lambda$ we can use the model of Sec.~\ref{sec:model} to estimate that at LHC at $y=0$ $Q_\mathrm{geom}=2.7$~GeV, and at $y=2$ $Q_\mathrm{geom}=5.0$~GeV. On the other hand, the original observation of \cite{Stasto:2000er} as well as the parameterization of \cite{Kharzeev:2004yx} seems to imply that the geometric scaling may still be a reasonable approximation at even higher momenta, provided that $x$ is small enough.  In order to illustrate dependence 
of $R_{pA}$ on $m_\bot$ we calculated it for four quark flavors shown in \fig{fig4}.
At masses of the order of  $m_t$ and at rapidity $y=2$ at LHC we have  $x\simeq 0.01$, so that $x$ is indeed small enough for the dipole model to be applicable. The amount of suppression of $R_{pA}$ depends on the anomalous dimension $\gamma$ as, roughly $R_{pA}\sim 1/N_\mathrm{coll}^{1-\gamma}$, where $\gamma$ is a function of $y$ and $m_\bot$, see \eq{gamma} and \eq{xii}.
Therefore, by measuring the $m_\bot$ and $y$ dependence of $R_{pA}$ one can determine the anomalous dimension $\gamma$ and map the transition region between the geometric scaling and the collinear factorization. The failure of the geometric scaling will manifest itself by rapid approach of the $R_{pA}$ to unity (modulo DGLAP corrections) at transverse mass $m_\bot\gtrsim Q_\mathrm{geom}$.

%%%%
%%%%%

\vskip0.3cm
{\bf Acknowledgments.}
I am grateful to Javier Albacete for showing me the results of his calculations of open heavy quark production; his results are in a qualitative agreement with \fig{fig2} and \fig{fig3}. I would like to thank RIKEN, BNL and the U.S. Department of Energy (Contract No. DE-AC02-98CH10886) for providing the facilities essential for the completion of this work.

\end{document}